\documentstyle[prl,floats,aps,twocolumn,epsf,graphicx]{revtex}

\newcommand{\beq}{\begin{equation}}
\newcommand{\eeq}{\end{equation}}
\newcommand{\beqr}{\begin{eqnarray} \nonumber}
\newcommand{\eeqr}{\end{eqnarray}}

\begin{document}
\twocolumn[\hsize\textwidth\columnwidth\hsize\csname
@twocolumnfalse\endcsname

\title{Survival Probabilities of History-Dependent Random Walks}
\author{Uri Keshet$^1$ and Shahar Hod$^2$ }
\address{$^1$Physics Faculty, Weizmann Institute, Rehovot 76100, Israel}
\address{}
\address{$^2$The Racah Institute of Physics, The Hebrew University, Jerusalem 91904, Israel}
\date{\today}
\maketitle

\begin{abstract}

\ \ \ We analyze the dynamics of random walks with long-term memory (binary chains with long-range correlations) in the presence of an absorbing boundary. An analytically solvable model is presented, in which a dynamical phase-transition occurs when the correlation strength parameter $\mu$ reaches a critical value $\mu_c$. For strong positive correlations, $\mu>\mu_c$, the survival probability is asymptotically finite, whereas for $\mu<\mu_c$ it decays as a power-law in time (chain length). 
\end{abstract}
\bigskip

]

Dynamical systems with long-range spatial or temporal correlations are attracting considerable interest across many disciplines, lending applications to physical, biological, social, and economical sciences (see, e.g., \cite{Man,Kan1,Sta,Pro,Usa,Yan,HodKeshet} and references therein). Such systems are often analyzed using random walks, a fundamental concept of statistical physics. Random walks lend applications to numerous scientific fields (see, e.g., \cite{BaNi,Kam,FeFrSo,Wei,AvHa,DiDa} and references therein). In particular, random walks in the presence of absorbing traps (boundaries) have been studied in recent years as models for various systems such as absorbing-state phase transitions \cite{MaDi,Hin}, polymer adsorption \cite{BeLo}, granular segregation \cite{FaFu}, the spreading of an epidemic \cite{GrChRo}, and in the context of complex adaptive systems \cite{HodNak1,Hod}. In this work we analytically study random walks with an absorbing boundary, in which the jump probability is history-dependent, resulting in long-range correlations. 

The statistical properties of data such as DNA strings, written texts and financial data (e.g., stock market quotes), are known to significantly deviate from those of purely random sequences \cite{Kan,Sch,HodKeshet}. Such systems may be studied by mapping them onto a correlated sequence of symbols. Although the nature of the resulting sequence may depend upon the choice of mapping (see, e.g., \cite{Narasimhan}), the essential statistical properties of the original system are often preserved. By choosing a mapping of these systems onto two symbols \cite{Usa}, the problem is reduced to the exploration of correlated binary chains, which are equivalent to one dimensional random walks with a constant step size. These binary chains have long-range correlations and often exhibit a super-diffusive nature, in which the variance grows asymptotically faster than the string length. 

The preceding discussion motivates a study of random walks with a history-dependent jump probability, both with and without an absorbing boundary, as means of facilitating our understanding of systems with long-range correlation. In Ref. \cite{HodKeshet} we presented a model for a history-dependent random walk. Although simple, this model features a dynamical phase transition between normal diffusion and super-diffusion as the correlation strength parameter reaches a critical value. In this work we analyze random walks with long-range correlations in the presence of an absorbing boundary. 

We begin by introducing a simple model that incorporates long-range correlations into an otherwise random sequence. Consider a discrete string of binary symbols, $a_i\in\{0,1\}$, in which the conditional probability of a given symbol (say, zero) occurring at the position $L+1$ is history-dependent, and is given by 
\begin{equation}\label{Eq1}
p(k,L)={1 \over 2}\Big(1-\mu {{L-2k} \over {L+L_0}}\Big)\  ,
\end{equation}
where $k$ is the number of such symbols (zeros) appearing in the preceding $L$ bits. 
The correlation parameter $\mu$, where $-1< \mu < 1$, determines the strength of correlations in the system. The persistence condition $\mu>0$ implies that a given symbol in the preceding sequence promotes the birth of a new identical symbol. In the anti-persistence regime $\mu<0$, on the other hand, each symbol inhibits the appearance of a new identical symbol. 
The parameter $L_0>1$ is a constant transient time. For $L \ll L_0$ the sequence is approximately 
random (uncorrelated), whereas for $L \gg L_0$ the effect of correlations takes over \cite{Note1}. 

In this model, the conditional probability $p(k,L;\mu,L_0)$ depends on the number of zeros (or unities) in the preceding bits, and is independent of their arrangement. This allows one to obtain an {\it analytical} description of the system's dynamical behavior. As demonstrated in \cite{HodKeshet}, this two-parameter model provides a good description of the observed statistical properties of various systems such as coarse-grained DNA strings, written texts, and financial data, when mapped onto a binary chain.

The probability $P(k,L+1)$ of finding $k$ identical symbols (say, zeroes) 
in a sequence of length $L+1$ follows from the evolution equation

\begin{eqnarray}\label{Eq2}
P(k,L+1) & = & p(k-1,L)P(k-1,L)  \nonumber \\
& & + [1-p(k,L)]P(k,L) \  .
\end{eqnarray} 
Crossing to the continuous limit, one obtains the 
Fokker-Planck diffusion equation for the correlated process,

\begin{equation}\label{Eq3}
{{\partial P} \over {\partial L}}={1 \over 2} {{{\partial^2 P} \over {\partial x}^2}}
-{{\mu} \over {L+L_0}}{{\partial(xP)} \over \partial x}\  ,
\end{equation}
where $x \equiv 2k-L$ is the distance from the origin in the corresponding random walk, and we have neglected high order terms which are irrelevant for $1\ll L_0 \ll L$. Solutions of Eq.~(\ref{Eq3}) under the initial condition $P(x,L=0)=\delta(x)$ were given in \cite{HodKeshet}.

We introduce an absorbing boundary at $x=0$, by imposing the boundary condition $P(x\leq0,L>0)=0$. The evolution equation (\ref{Eq3}) along with this boundary condition and the initial condition $P(x,L=0)=\delta(x)$, has a solution in the 
form 

\begin{equation}\label{Eq4}
P(x,L) \propto {{(L+L_0)}^{\mu} \over {D(L)^{3/2}}}\, x \Theta(x) \exp\Big[-{{x^2} \over {2D(L)}}\Big]\  ,
\end{equation}
where $\Theta$ is the Heaviside step function, and $D(L)$ is given by

\begin{equation}\label{Eq5}
D(L;\mu,L_0) \equiv {{L+L_0} \over {1-2\mu}}\Big[1-{\Big({{L_0} \over {L+L_0}}\Big)}^{1-2\mu}\Big]\  .
\end{equation} 
Equation (\ref{Eq5}) breaks down in the special case $\mu=\mu_c$, where $\mu_c\equiv 1/2$. In this case, $D(L)$ is given by 

\begin{equation}\label{Eq6}
D(L;\mu=\mu_c,L_0) \equiv (L+L_0)\ln\Big({{L+L_0} \over {L_0}}\Big)\  .
\end{equation}

The first two moments of the distribution function of the survived walkers are given by $\langle x \rangle^2 = \pi D(L)/2$ and $\langle x^2 \rangle = 2D(L)$. The variance of the probability distribution $P(x,L)$ thus equals

\begin{equation}\label{Eq8}
V(L;\mu,L_0)={{4-\pi} \over 2}D(L)\  .
\end{equation}
This result implies that for $\mu < \mu_c$, the asymptotic variance scales linearly with the string length, whereas for $\mu>\mu_c$ it scales as $V \propto L^{2\mu}$. Hence, a history-dependent sequence with strong positive correlations ($\mu > \mu_c$) is characterized by a super-diffusion phase in which the variance grows asymptotically faster than $L$, both without \cite{HodKeshet} and with an absorbing boundary.

The survival probability $S(L) \equiv \int_0^{\infty} P(x,t)dx$ of the walkers is given, for $\mu \neq \mu_c$, by 
\begin{equation} \label{eq:surv_neq_muc}
S(L; \mu\neq \mu_c) \propto \left| {\left({L+L_0 \over L_0}\right)^{1-2\mu} -1 }\right|^{-1/2} \, ,
\end{equation}
whereas for $\mu=\mu_c$ we find 
\begin{equation} \label{eq:surv_eq_muc}
S(L; \mu_c)\propto \ln^{-{1 \over 2}} \left({{L+L_0} \over L_0}\right) \, .
\end{equation}
%
The survival probability thus changes its asymptotic ($L \gg L_0$) behavior at the phase transition value $\mu_c=1/2$, and one identifies three qualitatively different regimes,

\begin{equation}\label{Eq11}
S(L \gg L_0) \propto \cases{ 
L^{-{1 \over 2}+\mu} & $\mu<\mu_c$\  ; \cr
\ln^{-{1 \over 2}} (L/L_0) & $\mu=\mu_c$\  ; \cr
const. & $\mu>\mu_c$\  . \cr }
\end{equation}

The normalization of the survival probability is sensitive to the discrete initial conditions. Since the chain is nearly random for $L\ll L_0$, the normalization may be approximated by equating S(L), for $1 \ll L \ll L_0$, to the survival probability of a purely random, continuous walk. We thus find 
\begin{equation} \label{eq:surv_norm}
S(1\ll L \ll L_0) \simeq \mbox{erf} \left(|x_0|/\sqrt{2L}\right) \, ,
\end{equation}
where $x_0$ is the distance between the absorbing boundary and the origin. 

\begin{figure}[tbh]
\centerline{\epsfxsize=9cm \epsfbox{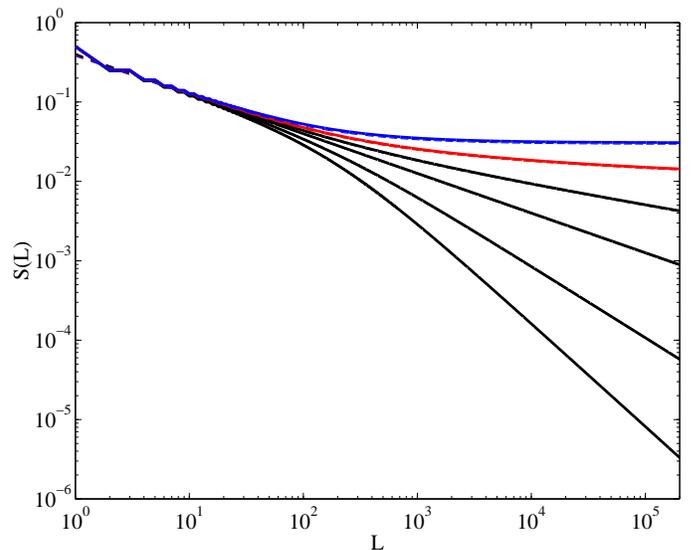}} 
\caption{Survival probability $S(L)$ as a function of sequence length $L$. 
Shown are numerical (solid lines) and analytical (dashed lines) results for $\mu=-0.8$, $-0.4$, $0$, $0.25$, $0.5$, and $0.8$ (from bottom to top). Here, $x_0=0$ (half the walkers survive the first step) and $L_0=100$. The analytical results agree with the numerical ones to better than $1\%$. 
}
\label{Fig1}
\end{figure}

In order to confirm the analytical results, we perform numerical simulations of (discrete) binary sequences. Figure \ref{Fig1} displays the resulting survival probability $S(L)$ of correlated strings with various values of the correlation parameter $\mu$. 
We find an excellent agreement between the analytical results [Eqs. (\ref{eq:surv_neq_muc}), (\ref{eq:surv_eq_muc}) and (\ref{eq:surv_norm})] and the numerical ones.

\begin{figure}[tbh]
\centerline{\epsfxsize=9cm \epsfbox{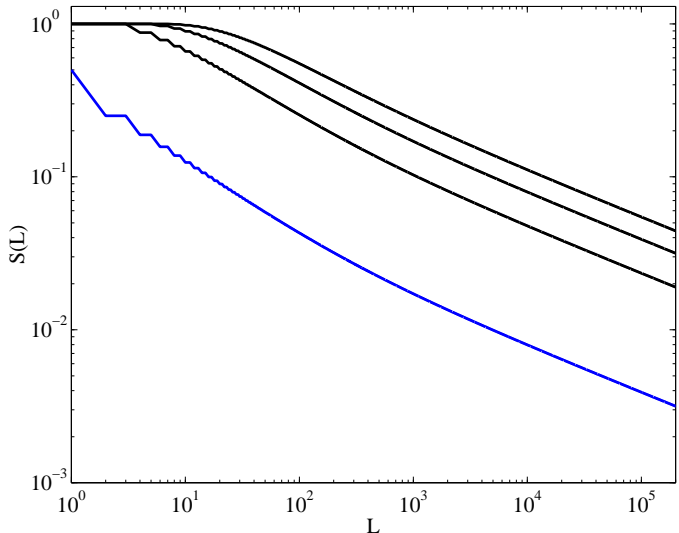}} 
\caption{Survival probability $S$ as a function of string length $L$ for various boundary locations $x_0$. We present results for $x_0=0, -2, -4$, and $-6$ (from bottom to top), with $\mu=0.2$ and $L_0=100$. 
}
\label{Fig2}
\end{figure}

The preceding results, starting from the distribution function in Eq. (\ref{Eq4}), are valid only when the absorbing boundary is placed at the origin. One would like to generalize the results for boundaries located at arbitrary locations $x_0$, under the boundary condition $P(x\leq x_0,L>0)=0$. Note that the diffusion equation (\ref{Eq3}) is {\it not} invariant under the translation $x \to x+d$, where $d$ is a constant. The generalization of our solution for $x_0 \neq 0$ is therefore non-trivial. 
Nevertheless, we have verified numerically that the asymptotic behavior of the survival probability given by Eq. (\ref{Eq11}) remains valid for arbitrary values of $x_0$, in all three regimes. The similar asymptotic behavior for different choices of $x_0$ is demonstrated numerically in Fig. \ref{Fig2}.

The above results can be readily generalized for a biased random walk with a moving boundary. For example, for the biased jump probability 
\begin{equation} 
p(k,L)={1 \over 2} \left( 1+q-\mu {L-2k \over {L+L_0}} \right) \,
\end{equation}
the above results will hold if we apply the transformation $x\rightarrow x-x_c(L)$, with 
\begin{equation}
x_c(L) \equiv q {L+L_0 \over {1-\mu}} \left[1-\left({L_0 \over {L+L_0}}\right)^{1-\mu} \right] \, .
\end{equation}

In summary, in this work we have analyzed the dynamics of random walks in which 
the jump probabilities are history-dependent, in the presence of an absorbing boundary. Using an analytically solvable model, we identify a dynamical phase transition characterizing the system's global behavior. 
For small values of the correlation strength ($\mu<\mu_c$) the 
survival probability decays as $S \propto L^{-(\mu_c-\mu)}$, whereas for $\mu > \mu_c$ the system is characterized by {\it finite} asymptotic survival probabilities. 

\bigskip
\noindent
{\bf ACKNOWLEDGMENTS}
\bigskip

The research of SH was supported by G.I.F. Foundation.

\end{document}